\documentclass[pra,aps,twocolumn,epsfig,superscriptaddress,showpacs]{revtex4}
\usepackage{amssymb}
\usepackage{amsmath}
\usepackage{epsfig}
\usepackage{amsfonts}
\usepackage{color}
\usepackage{float}
\usepackage{latexsym}
\usepackage{mathrsfs}
\usepackage{feynmf}
\def\QED{\leavevmode\unskip\penalty9999 \hbox{}\nobreak\hfill
     \quad\hbox{\leavevmode  \hbox to.77778em{%
               \hfil\vrule   \vbox to.675em%
               {\hrule width.6em\vfil\hrule}\vrule\hfil}}
     \par\vskip3pt}
\def\qed{\leavevmode\unskip\penalty9999 \hbox{}\nobreak\hfill
     \quad\hbox{\leavevmode  \hbox to.77778em{%
               \hfil\vrule   \vbox to.675em%
               {\hrule width.6em\vfil\hrule}\vrule\hfil}}
\par\vskip3pt}
\def\ibb #1{\leavevmode\hbox{\kern.3em\vrule
     height 1.5ex depth -.1ex width .4pt\kern-.3em\rm#1}}

\newcommand{\be}{\begin{equation}}
\newcommand{\ee}{\end{equation}}
\newcommand{\ba}{\begin{array}}
\newcommand{\ea}{\end{array}}
\newcommand{\bqa}{\begin{eqnarray}}
\newcommand{\eqa}{\end{eqnarray}}

\newcommand{\tr}{\mbox{Tr}}
\newcommand{\bra}[1]{\ensuremath{\langle #1 |}}
\newcommand{\ket}[1]{\ensuremath{| #1 \rangle}}

\begin{document}

\author{ZhiHao Ma}
\email[Email:]{ma9452316@gmail.com}
\affiliation{Department of Mathematics, Shanghai Jiaotong
University, Shanghai, 200240, P.R.China}

\author{Xiao-Dong Zhang}
\email[Email:]{xiaodong@sjtu.edu.cn} \affiliation{Department of
Mathematics, Shanghai Jiaotong University, Shanghai, 200240,
P.R.China}

\date{\today}

\begin{abstract}

A new entanglement measure, which is called D-concurrence, is
proposed. Then the upper and lower bounds for D-concurrence are
obtained and the relationship between D-concurrence and the usual
concurrence of Wootters was established.  In addition,
 comparing with the usual concurrence, D-concurrence has some special merits.
\end{abstract}

\title{A new entanglement measure: D-concurrence}
 \maketitle

\section{Introduction}

Quantum entanglement is the key resource in quantum information processing and quantum computation\cite{Horodecki09, Nielsen00, Guhne09}.

A mix state $\rho$ is called {\it separable} if it can be written as
a convex combination of tensor product states \cite{Werner89}
\begin{equation}\label{separable}
\rho = \sum_j p_j \rho_j^{(A)}\otimes\rho_j^{(B)},
\end{equation}
otherwise it is {\it entangled} or {\it inseparable}.

Then a question arises: How to detect whether a state  is entangled
or not? If $\rho$ is  entangled, how to quantify the degree of its
entanglement? To answer these two questions, we need to introduce some measures of entanglement
to quantify the degree of entanglement. The entanglement measure for a state is zero iff the state is separable, and the bigger is the entanglement measure,  then more entanglement is the state. One of the most famous
measures of entanglement is the concurrence \cite{Wootters98} of
two-qubit system. The concurrence of a pure two-qubit state $\psi$
is given by
\begin{eqnarray} \label{ConP}
C(| \psi \rangle) =\sqrt{2[1- \tr {\rho_A}^2]}= \sqrt{2[1- \tr
{\rho_B}^2]},
\end{eqnarray}
where $\rho_A= \tr_B | \psi \rangle \langle \psi |$ is the partial
trace of $|\psi \rangle \langle \psi |$ over subsystem $B$, and
$\rho_B$ has a similar meaning. For a mixed state, the concurrence
is defined by the convex roof method, that is, as the average concurrence of the pure states of the
decomposition, minimized over all decompositions of $\rho =
\sum_{j}p_{j} | \psi_j \rangle  \langle \psi_j |$,
\begin{eqnarray}\label{ConM}
C(\rho)= \min \sum_{j}p_{j}C(| \psi_j \rangle).
\end{eqnarray}
It is first discovered in \cite{Wootters98} a simple way to quantify
the concurrence of two-qubit mixed state $\rho_{AB}$
\begin{eqnarray}
C=\max\{\lambda_1-\lambda_2-\lambda_3-\lambda_4,0\},
\end{eqnarray}
where $\lambda_i$ is the square root of eigenvalues of
$\rho^{AB}\cdot(\sigma_y\otimes\sigma_y)\cdot(\rho^{AB})^*\cdot
(\sigma_y\otimes\sigma_y)$ in decreasing order. The definition is
also work for pure-state case of two qubits, in this case it
possesses a simpler form
\begin{eqnarray}
\label{pureC2}C=2\sqrt{\det(\rho_A)}=2\sqrt{\det(\rho_B)},
\end{eqnarray}
where $\rho_A$ and $\rho_B$ are reduced density matrices obtained
from the pure state $\rho_{AB}$ by tracing out the other particle,
and $\det $ is the determinant function of the matrix.

For a general high dimension pure bipartite state $ \vert\Psi\rangle $, $ \vert\Psi\rangle\in\mathcal{H}_{A}\otimes \mathcal{H}_{B}\ $, concurrence is defined as~\cite{Mintert05}:
\begin{equation}
C\left( \Psi\right)=\sqrt{2 [\langle\Psi\vert\Psi\rangle^{2}-\tr\rho_{i}^{2} ]}\,,
\label{1}
\end{equation}
where $ \rho_{i} $ is the reduced density operator obtained by tracing over either
subsystems A or B. It is clear that $ C(\Psi)=0 $ if and only if $ \vert\Psi\rangle $ is a product state, i.e.
$ \vert\Psi\rangle= \vert\Psi_{A}\rangle\otimes \vert\Psi_{B}\rangle $.

Interestingly, $C(\Psi)$ can be observed through a small number
of projective measurement on a \textit{twofold copy}
$\ket{\Psi}\otimes\ket{\Psi}$ of
$\ket{\Psi}$ ~\cite{Mintert04,Mintert05,Mintert05PR,Mintert06,Mintert07, Mintert07A,Mintert08}:
\begin{equation}
C\left( \Psi\right)=\sqrt{\bra{\Psi}\otimes\bra{\Psi}\ A\ \ket{\Psi}\otimes\ket{\Psi}}\,, \\
{\cal A}=4 P_{-}^{A}\otimes P_{-}^{B}\,,\qquad\qquad
\label{2}
\end{equation}
where $ P_{-}^{A} $ ($ P_{-}^{B} $) is the projector onto the antisymmetric subspace of $ \mathcal{H}_{A}\otimes \mathcal{H}_{A}\ $ ($ \mathcal{H}_{B}\otimes \mathcal{H}_{B}\ $).

For mixed states the concurrence is defined by the convex roof method ~\cite{Mintert07}:
\begin{align}
C(\rho)= \min \sum_{i}p_{i}C( \Psi_i )\,, \qquad\qquad\notag \\
\rho =\sum_{i}p_{i} \vert\Psi_{i}\rangle\langle\Psi_{i}\vert\,, \qquad p_{i}\geq 0\,, \qquad\sum_{i}p_{i} =1\,,
\label{4}
\end{align}
where the minimum is taken over all decompositions of $\rho $ into pure states $ \vert\Psi_{i}\rangle $ .

Since the concurrence for high dimension mix state is difficult to
calculate, it is a urgent task to find bound for concurrence. Until
now, only a few bounds for concurrence have been
obtained~\cite{Fei04,Fei05,Fei06,Fei08,Fei09,Sargolzahi09,Zhang08,Guhne08,Guhne08,Ma09,Breuer06,Vicente0708,Augusiak09}.

In 2007,  Mintert and  Buchleitner gave a lower bound for
concurrence as\cite{Mintert07}:
\begin{eqnarray} \label{ineq}
C^{2}(\rho) \ge 2 \bigr[ \tr\rho^{2}-\tr{\rho_A}^{2} \bigr],
\end{eqnarray}
where $C(\rho)$ is the concurrence for \textit{arbitrary} states,
taking the definitions in Eq. (7) and Eq. (8).

Recently,  a lower bound for concurrence was discovered
(\cite{Zhang08,Ma09}):
\begin{eqnarray} \label{ineq}
C^{2}(\rho) \leq 2 \bigr[1-\tr{\rho^A}^{2} \bigr],
\end{eqnarray}

In this paper, we will introduce a new entanglement measure, which
is called  D-concurrence. The upper and lower bounds for
D-concurrence are also discussed.

\section{Main results}

It is known that a pure state $| \psi \rangle$ is separable if and
only if its two reduced density matrices $\rho_{A},\rho_{B}$ are all
pure state. So for pure states $\rho$, the D-concurrence is defined
as $D(\rho):=\sqrt{\det(I-\rho_A)}$, where $\det$ is the determinant
function of a matrix.

For a mixed state,  D-concurrence is defined by the convex roof, that is, defined as the average
D-concurrence of the pure states of the decomposition, minimized
over all decompositions of $\rho = \sum_{j}p_{j} | \psi_j \rangle
\langle \psi_j |$,
\begin{eqnarray}\label{ConM}
D(\rho):= \min \sum_{j}p_{j}D(| \psi_j \rangle).
\end{eqnarray}
A decomposition whose  convex combination reaches the minimum is
called an optimal one.

Comparing with usual concurrence, D-concurrence has some advantages:

First, it is defined by determinant $\det(I-\rho_{A})$, which is a
first order function of $\rho_{A}$, while concurrence is defined by
$\tr{\rho_A}^{2}$, which is a two-order function of $\rho_{A}$, and
it is known that first order is easy to handle in some sense.

Interestingly,  there is a deep connection between concurrence and D-concurrence:

{\bf Proposition 1.} For the case of two-qubits state, $D(\rho)=\frac{1}{2}C(\rho)$.

{\bf Proof.} Easily.

We can get an upper bound of D-concurrence, that is:

{\bf Theorem 1.} For any bi-particle states $\rho$, we have
\begin{eqnarray} \label{ineq}
D^{2}(\rho) \leq  \bigr[ \det(I-\rho_{A})\bigr],
\end{eqnarray}

{\bf Proof.} Suppose $\rho$ has a decomposition as $\rho=\sum_{i}p_{i}|\psi_i
\rangle  \langle \psi_i |$. Then
\begin{eqnarray}
[D(\rho)]^{2}&=&[\inf
\sum_{i}p_{i}D(| \psi_i \rangle  \langle \psi_i |)]^{2}\nonumber\\
&\leq&\inf
\sum_{i}[\sqrt{p_{i}}D(| \psi_i \rangle  \langle \psi_i |)]^{2}\cdot\sum_{i}(\sqrt{p_{i}})^{2}\nonumber\\
&=&\inf\sum_{i}p_{i}\det(I-(\psi_i)_{A})\nonumber\\
&\leq& \det(I-\rho_{A})\nonumber\\\nonumber
\end{eqnarray}
where $\rho_{A}:=\mathrm{Tr}_{B}\rho$ is the reduced density
matrices of  $\rho$, and
$(\psi_i)_{A}:=\mathrm{Tr}_{B}|\psi_{i}\rangle\langle\psi_{i}|$ be
the reduced density matrix of $|\psi_i \rangle  \langle \psi_i |$.
The first inequality holds by applying the Cauchy-Schwarz
inequality, while the second inequality holds due to the following
result: Assume $A, B$ are two Hermitian matrices, $A\geq 0, B\geq 0$, i.e.,
semi-positive definite. Then \begin{eqnarray}\det(A+B)\geq
\det(A)+\det(B)\end{eqnarray} The proof of inequality (13) is referred to \cite{Hua}. This finishes our proof.

{\bf Note.} From numerical experiment, one can see that (12) is
better than inequality (10), that is, the new bound is more closer to the real value of concurrence.

How about lower bound? We first prove the following:

{\bf Theorem 2.} If $\rho_{AB}$ is a separable state, then the following holds:
\begin{eqnarray} \label{ineq}
 \det(I-\rho_{A})- \det(I-\rho)\leq 0,
\end{eqnarray}

{\bf Proof.} From \cite{Nielsen01}, we get that if
$\rho _{AB}$ is separable, then
\begin{equation} \label{eq:majorizationA}
\lambda (\rho _{AB})\prec \lambda (\rho _{A}),
\end{equation}
and
\begin{equation} \label{eq:majorizationB}
\lambda (\rho _{AB})\prec \lambda (\rho _{B}),
\end{equation}
where $\lambda (\rho _{AB})$ is a vector of eigenvalues of $\rho
_{AB}$; $\lambda (\rho _{A}) $ and $\lambda (\rho _{B})$ are defined
similarly. The relation $x\prec y$
between $n$-dimension vectors $x$ and $y$, which reads ``$x$ is majorized by $%
y $'', means that
\begin{equation} \label{eq:majdef1}
\sum_{i=1}^{k}x_{i}^{\downarrow }\leq
\sum_{i=1}^{k}y_{i}^{\downarrow }\qquad (1\leq k\leq n-1),
\end{equation}
and
\begin{equation} \label{eq:majdef2}
\sum_{i=1}^{n}x_{i}^{\downarrow }=\sum_{i=1}^{n}y_{i}^{\downarrow },
\end{equation}
where $x_{i}^{\downarrow }$ $(1\leq i\leq n)$ are components of vector $%
x $ rearranged in decreasing order ($x_{1}^{\downarrow }\geq
x_{2}^{\downarrow }\geq \cdots \geq x_{n}^{\downarrow }$);
$y_{i}^{\downarrow }$ $(1\leq i\leq n)$ are defined similarly. If
the dimensions of $x$ and $y$ are different, the smaller vector is
enlarged by appending extra zeros to equalize their dimensions.

We know that for vectors $x,y$, $x\prec y$ if and only if $\sum_{i}f(x_{i})\leq \sum_{i}f(y_{i})$ for all continuous convex function $f:R\mapsto R$. Note that $x\prec y$ then $I-y \prec I-x$, here $I:=(1,1,...1)$ is the unit vector.

Denote $x:=(x_{1},x_{2},...x_{n})$ as vector of eigenvalues of $\rho
_{AB}$, $y:=(y_{1},y_{2},...y_{n})$ as vector of eigenvalues of $\rho
_{A}$ (appending extra zeros to equalize the dimension of $x$), then we get that
$\det(I-\rho_{A})=\prod\limits_{i}(1-y_{i})$, $\det(I-\rho_{AB})=\prod\limits_{i}(1-x_{i})$.

Now we will discuss the following four possible cases:

(1). If $x:=(1,0,...0)$, $y\neq (1,0,...0)$, then $x\prec y$ can not happen.

(2). If $x:=(1,0,...0)$, $y= (1,0,...0)$, then $\det(I-\rho_{A}) =\det(I-\rho_{AB})$,   $\det(I-\rho_{A})- \det(I-\rho)\leq 0$.

(3). If $x\neq (1,0,...0)$, $y= (1,0,...0)$, then $\det(I-\rho_{A})- \det(I-\rho)< 0$.

(4). If $x\neq (1,0,...0)$, $y\neq (1,0,...0)$, then all $x_{i}, y_{i}$ satisfying that $0\leq x_{i}<1, 0\leq y_{i}<1$,
then we get that  $\sum_{i}f(1-y_{i})\leq \sum_{i}f(1-x_{i})$ for all continuous  convex function $f:R\mapsto R$, we choose the function $f(t):=-\log t$, then we get $\sum_{i}-\log(1-y_{i})\leq \sum_{i}-\log(1-x_{i})$.

 All the above show that $\det(I-\rho_{A})- \det(I-\rho)\leq 0$. Theorem is proved.

Numerical experiments show the following is true:

{\bf Proposition 2.} For a general state $\rho$, D-concurrence has a lower bound as:
\begin{eqnarray} \label{ineq}
D^{2}(\rho) \ge  \bigr[ \det(I-\rho_{A})- \det(I-\rho)\bigr].
\end{eqnarray}

Comparing the inequality (19) with the inequality (9), from numerical experiments,  we can find that
our result is better than (9).

{\bf Example 1.} Consider the werner state,
\begin{eqnarray}\rho_{f}:=\frac{1}{N^3-N}[(N-f)I+(Nf-1)P],\end{eqnarray}
where $P:=\sum | ij \rangle \langle ji |$ is the swap operator, $f$
is a constant number, $-1\leq f\leq 1$. We know that
$C^2(\rho_{f})=f^2$. Take $f=-\frac{1}{2}$, then
$C^2(\rho_{f})=0.25$, while $4\bigr[\det(I-\rho_{A})-
\det(I-\rho)\bigr]=0.2297, 2 \bigr[ \tr\rho^{2}-\tr{\rho_A}^{2}
\bigr]=0.1667$. Our bound is closer to the real concurrence!

\section{Conclusions}
In this paper, we define a new entanglement measure, called
D-concurrence. It is seen that D-concurrence has deep connection
with the usual concurrence, and also has its own advantages. We then
obtain the lower and upper bounds for D-concurrence. Comparing with
the bounds for usual concurrence, our bound is  closer to the real
values of concurrence.

Also,  we leave some open questions, and will study in future.
One of them is the following: What is the physical interpretation of $D-$concurrence?
 We know that the result of \cite{Mintert07} is very interesting,
 because we can experiment detect it. How about $D-$concurrence?

{\bf ACKNOWLEDGMENTS} The first author would thanks Prof.Man-Duen Choi for valuable discussion and help.
This work is supported by NSF of China(NSFC 10901103), and the New teacher
Foundation of Ministry of Education of P.R.China (Grant No.
20070248087), partially supported by a grant of science and
technology commission of Shanghai Municipality (STCSM, No.
09XD1402500). XiaoDong Zhang is supported by National Natural Science Foundation of China
(No.10531070 No.10971137), National Basic Research Program of China 973 Program (No.2006CB805900),
a grant of Science and Technology Commission of Shanghai Municipality (STCSM, No.09XD1402500).

\end{document}